\documentclass[conference,10pt,final]{IEEEtran}

\usepackage[T1]{fontenc}
\usepackage{cite}
\usepackage{url}
\usepackage{amsmath}
\usepackage{amssymb}
\usepackage[cmintegrals]{newtxmath}
\usepackage[labelformat=simple]{subcaption}

\usepackage{dblfloatfix}
\usepackage{graphicx}
\usepackage{epstopdf}
\usepackage{xcolor}
\usepackage{mathtools,amssymb,lipsum}
\usepackage[detect-weight=true, binary-units=true]{siunitx}
\usepackage[inline]{enumitem}
\usepackage{cuted}
\usepackage[figure]{algorithm2e} 
\usepackage{booktabs}

\usepackage{float}

\IEEEoverridecommandlockouts
\IEEEpubid{\makebox[\columnwidth]{978-1-5386-1478-5/18/\$31.00 \copyright~Copyright 2018
		IEEE \hfill} \hspace{\columnsep}\makebox[\columnwidth]{ }}

\begin{document}

\title{5G Island for Network Resilience and Autonomous Failsafe Operations}

\author{
	\IEEEauthorblockN{
		Bin Han\IEEEauthorrefmark{1}, Marcos Rates Crippa\IEEEauthorrefmark{1} and Hans D. Schotten\IEEEauthorrefmark{1}
	}
	\IEEEauthorblockA{\IEEEauthorrefmark{1}Institute of Wireless Communication (WiCon), Technische Universit\"at Kaiserslautern\\ Paul-Ehrlich-Stra{\ss}e 11, 67663 Kaiserslautern, Germany\\
		Emails: \{binhan,crippa,schotten\}@eit.uni-kl.de}
}


%


\maketitle

\begin{abstract}
The resilience of 5G networks can be strongly challenged by central cloud virtual network function (VNF) outages, which can be cause by server and backhaul connection errors. This paper proposes a context-aware approach to migrate VNF from central cloud to local edge cloud, in order to improve the network resilience with minimized VNF migration cost.
\end{abstract}

\IEEEpeerreviewmaketitle

\section{Introduction}
The envisaged scenario of ultra-reliable low-latency communications (URLLC) in the Fifth Generation (5G) cellular networks calls for a $99.999\%$ transmission correctness within \SI{1}{\milli\second} \cite{3gpp2017study}, in order to support safety critical applications such like vehicular communications\cite{ji2017feasibility}
. In these use cases, network resilience and autonomous failsafe operations against service outages become extremely critical.

Existing techniques to achieve high resilience in software defined network (SDN) include state management, virtual network function (VNF) migration and rollback recovery\cite{han2017resiliency}, which all require some kind of VNF redundancy. 
For VNFs running on central cloud servers, the VNF migration approach is more effective, as it is able to hand with the outages caused by degraded backhaul connection by placing redundant instance copies of the central cloud VNFs in the local edge cloud and periodically updating them. However, a considerable extra cost will be generated as a trade-off.

In this paper, we propose a context-aware central-to-edge (C2E) VNF migration framework, the \emph{``5G Island''} \textit{5G Island} (5GI), which makes use of context information about users and telco clouds, and cognitively prepare redundant instances of central cloud VNFs on edge cloud servers, to enable make-before-break VNF migration for better network resilience and autonomous failsafe operations with a reasonable cost.



\section{Central-to-Edge VNF Migration}\label{sec:c2e_vnf_migration}
For central cloud VNFs, an error can occur either at the central cloud server, or at the backhaul connection. In the latter case, neither state management nor rollback recovery efficiently helps, while a VNF migration from the central cloud to the local edge cloud does.
Central cloud VNF outages can be triggered by planned disconnections, cyber attacks and unintentional errors. Cyber attacks to paralyze network services are usually followed by attempts of illegal access to confidential information. Unintentional critical network infrastructure malfunctions can be caused by dangerous disasters and emergencies such like fire, explosion, earthquake and massive blackout. In both cases, autonomous failsafe operations shall be executed independently from the central cloud, where C2E VNF migration is required.


It costs not only data traffic to update the redundancies of central cloud VNFs in edge clouds, but also server resource to maintain them. Hence there is a trade-off between the network resilience and the operations expenditure (OPEX) when deploying the C2E VNF migration. Aiming to take a reasonable balance, we need an intelligent and predictive method that only migrates a central cloud VNF to edge cloud server, when the opportunity loss in case of central cloud VNF outage exceeds the cost of C2E VNF migration itself.

\section{5G Island as a Context-Aware C2E VNF Migration Framework}\label{sec:5g_island}
\IEEEpubidadjcol 
To cope with the demands discussed above, we come to the novel concept of a context-aware VNF migration framework, which we call the \emph{5G Island}. In this concept, every edge cloud periodically and selectively updates VNF redundancies from the central cloud to mitigate potential loss caused by central cloud VNF outages.

A central cloud VNF outage will lead to a loss of every user equipment (UE) that relies on this VNF, in case the VNF is not migrated to the local edge cloud. Given a certain central cloud VNF and a certain edge cloud, the total \emph{opportunity cost} of outage for the VNF in the local edge coud is
\begin{equation}\label{equ:opp_cost}
	c_\text{o}=\sum\limits_{u\in\mathcal{U}}\text{E}\{t_{\text{o},u}\}\cdot l,
\end{equation}
where $\mathcal{U}$ is the set of all UEs that may be served by the local edge cloud in the next updating period, $\text{E}\{t_{\text{o},u}\}$ is the expected time $t$ that the user device $u$ suffers from an outage of the given VNF in the next updating period $T$, and $l$ denotes the outage loss of each served UE in unit time. As aforementioned, it generates a certain cost to update and locally maintain the VNF redundancy, which we call the \emph{migration cost}, denoted by $c_\text{m}$. To minimize the overall expected cost, for every VNF the edge cloud should decide:
\begin{equation}
	\begin{cases}
		\text{To synchronize the redundancy} & c_\text{m}\le c_\text{o}\\
		\text{Not to synchronize the redundancy} & c_\text{m}> c_\text{o}\\
	\end{cases}
\end{equation}

While both $c_\text{m}$ and $l$ are generally static and can be evaluated by the mobile network operator (MNO) straight forwardly, $\text{E}\{t_{\text{o},u}\}$ can be dynamic and depends on characteristics of users and the environment. Assuming that the period of redundancy updating is $T$, 
\begin{equation}\label{equ:outage_model}
	\text{E}\{t_{\text{o},u}\}\approx p_o\eta_u\int_{0}^{T}f_{\text{arr},u}(t)\int_{0}^{T-t}f_{\text{stay},u}(\tau)\tau\text{d}\tau\text{d}t
\end{equation}
where $p_\text{o}$ is the predicted VNF outage rate in the next period $T$, $\eta_u$ is the VNF duty rate for the UE $u$, $f_{\text{arr},u}(t)$ denotes the probability density that $u$ arrives in the coverage of local edge cloud after $t$, and $f_{\text{stay},u}(\tau)$ is the probability density function of $u$'s stay time $\tau$ in the local area upon arrival.

It shall be noted that the 5GI differentiates from the similar concepts of Trust Zone and 5G Private Networks, see\cite{kochems2018ammcoa}.

\section{Context Provisioning and Opportunity Cost Estimation}\label{sec:provisioning_estimation}
Generally, there are two approaches to estimate the opportunity cost defined in (\ref{equ:opp_cost}), namely UE-driven and edge-cloud-driven estimations, respectively.

For stateful VNFs such like a virtualized Home Subscriber Server (HSS), where UE-relevant data have to be synchronized together with the function in the mitigation, the local edge cloud has to know the identifications of the UEs that are likely arriving. In this case, the set $\mathcal{U}$ of UEs that may arrive in the local edge cloud must be first estimated, so that $\eta_u$, $f_{\text{arr},u}$ and $f_{\text{stay},u}$ can be individually estimated for every $u\in\mathcal{U}$. Essential context information needed for such an estimation include the current UE positions, the UE mobilities and the physical traffic conditions in near areas. Besides, $p_\text{o}$ must be predicted with the context information of current VNF performances.

For stateless VNFs which has no dependency on the served UEs such as a gateway, all devices are homogeneously processed in the VNF migration. In this case, an edge-cloud-driven estimation can be applied to reduce the effort of provisioning and processing context information. Each edge cloud only needs to estimate the expected number of UEs served by the central cloud VNF per updating period, and the expected service time per UE. Instead of directly calling (\ref{equ:opp_cost}) and (\ref{equ:outage_model}), $c_\text{o}$ is here estimated as
\begin{equation}
	\hat{c}_\text{o}=\bar{N}\bar{\eta}p_\text{o}\int_{0}^Tf_{\text{stay}}(\tau)\text{d}\tau\text{d}t,
\end{equation}
where $\bar\eta$ is the average duty factor of the target VNF in local edge cloud, $\bar{N}$ is the average number of UEs relying on the target VNF and being served by the local edge cloud, $f_{\text{stay}}(\tau)$ is the probability density function of their local stay time. $\bar{\eta}$ can be obtained by the VNF Manager (VNFM) from historical data. Both $\bar{N}$ and $f_{\text{stay}}(\tau)$ depend on the current statistics of user mobility, which can be provisioned with assistance of the Access \& Mobility management Function (AMF). Similarly as for the UE-driven estimation, $p_\text{o}$ must be predicted from the current VNF performances.

\section{Simulation}\label{sec:simulations}
\begin{figure}[!h]
	\centering
	\includegraphics[width=.45\textwidth]{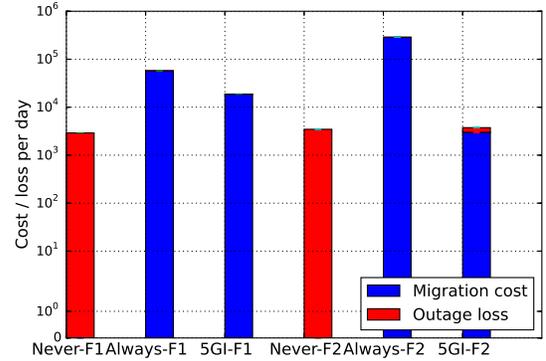}
	\caption{Simulated C2E migration cost and outage loss}
	\label{fig:cost_loss}
\end{figure}

Fig.\ref{fig:cost_loss} shows the simulated daily migration cost and outage loss of a central cloud VNF. The simulation scenario considered an edge cloud covering an \SI{4\pi}{\kilo\meter^2} urban area surrounded by suburban and rural areas. UEs of different mobility classes randomly move over these areas n result in an UE density of \SI{187.23}{\kilo\meter^{-2}} in the edge cloud coverage. Two different central cloud VNFs, F1 and F2, were specified to have independent random time-varying availabilities as Markov processes. For F1 there was $c_\text{m}=20Tl$ and for F2 $c_\text{m}=100Tl$. For both VNFs and any UE $u$ we set $\eta_u=1$. We tested the 5GI approach in comparison with the baselines of never/always updating the redundancy. The results shows clearly that our proposed approach effectively benefits the network resilience with reduced VNF migration cost.

\section{Conclusion}\label{sec:conclusion}
By estimating the opportunity VNF outage loss and comparing it to the VNF migration cost, our proposed intelligent context-aware C2E VNF migration framework, the \emph{5G Island}, improves the network resilience with a minimized total cost.

\section*{Acknowledgment}
This work has been supported by the EU Horizon-2020 project 5G-MoNArch (grant agreement number 761445). The authors would like to acknowledge the contributions of their colleagues. This information reflects the consortium's view, but the consortium is not liable for any use that may be made of any of the information contained therein.

\bibliographystyle{IEEEtran}
\bibliography{references}

\begin{thebibliography}{1}
\providecommand{\url}[1]{#1}
\csname url@samestyle\endcsname
\providecommand{\newblock}{\relax}
\providecommand{\bibinfo}[2]{#2}
\providecommand{\BIBentrySTDinterwordspacing}{\spaceskip=0pt\relax}
\providecommand{\BIBentryALTinterwordstretchfactor}{4}
\providecommand{\BIBentryALTinterwordspacing}{\spaceskip=\fontdimen2\font plus
\BIBentryALTinterwordstretchfactor\fontdimen3\font minus
  \fontdimen4\font\relax}
\providecommand{\BIBforeignlanguage}[2]{{%
\expandafter\ifx\csname l@#1\endcsname\relax
\typeout{** WARNING: IEEEtran.bst: No hyphenation pattern has been}%
\typeout{** loaded for the language `#1'. Using the pattern for}%
\typeout{** the default language instead.}%
\else
\language=\csname l@#1\endcsname
\fi
#2}}
\providecommand{\BIBdecl}{\relax}
\BIBdecl

\bibitem{3gpp2017study}
``{Study on Scenarios and Requirements for Next Generation Access Technologies
  (Rel-14)},'' Third Generation Partnership Project (3GPP), Tech. Rep. TR
  38.913, Aug. 2017.

\bibitem{ji2017feasibility}
L.~Ji, A.~Weinand, B.~Han, and H.~D. Schotten, ``{Feasibility Study of Enabling
  V2X Communications by LTE-Uu Radio Interface},'' in \emph{2017 IEEE/CIC
  International Conference on Communications in China (ICCC)}, Oct 2017, pp.
  1--6.

\bibitem{han2017resiliency}
B.~Han, V.~Gopalakrishnan, G.~Kathirvel, and A.~Shaikh, ``{On the Resiliency of
  Virtual Network Functions},'' \emph{IEEE Communications Magazine}, vol.~55,
  no.~7, pp. 152--157, 2017.

\bibitem{kochems2018ammcoa}
J.~Kochems and H.~D. Schotten, ``{AMMCOA-Nomadic 5G Private Networks},''
  \emph{arXiv preprint arXiv:1804.07665}, 2018.

\end{thebibliography}

\end{document}